\documentclass[12pt]{article}
\usepackage{epsfig}
\begin{document}
\begin{titlepage}
\begin{center}

{\Large Complexity of chaotic fields and standard model
parameters}

\vspace{1cm}

{\bf Christian Beck}

School of Mathematical Sciences, Queen Mary, University of
London, Mile End Road, London E1 4NS, UK

\end{center}

\abstract{In order to understand the parameters of the standard
model of electroweak and strong interactions (coupling constants,
masses, mixing angles) one needs to embed the standard model into
some larger theory that accounts for the observed values. This
means some additional sector is needed that fixes and stabilizes
the values of the fundamental constants of nature. In these
lecture notes we describe in non-technical terms how such a
sector can be constructed. Our additional sector is based on
rapidly fluctuating scalar fields that, although completely
deterministic, evolve in the strongest possible chaotic way and
exhibit complex behaviour. These chaotic fields generate
potentials for moduli fields, which ultimately fix the fundamental
parameters. The chaotic dynamics can be physically interpreted in
terms of vacuum fluctuations. These vacuum fluctuations are
different from those of QED and QCD but coupled with the same
moduli fields as QED and QCD are. The vacuum energy generated by
the chaotic fields underlies the currently observed dark energy
of the universe. Our theory correctly predicts the numerical
values of the electroweak and strong coupling constants using
a simple principle, the minimization of vacuum energy.
Implementing some additional discrete symmetry assumptions one
also obtains predictions for fermion masses, as well as a Higgs
mass prediction of 154 GeV.}

\end{titlepage}

\section{Introduction}

String theories predict an enourmous amount of possible vacua
after compactification, of the order $10^{120}$. In each of these
vacua the cosmological constant as well as the fundamental
constants of nature can have different values. One is led to the
so-called `landscape' picture \cite{susskind}. The landscape in a
sense represents the set of all possible blueprints of the
universe. To select the right vacuum, i.e. the one we observe
around us right now, often an anthropic point of view is chosen.
But is an anthropic principle really the last word and the
ultimate answer to all our questions?

A natural idea to avoid an anthropic selection principle would be
that the universe will not be left alone with its choice of
$10^{120}$ vacua but that it gets some help. This help should have
the form of an additional sector, a theory as yet not included in
the ordinary standard model, neither in ordinary string theories.
The additional sector should yield some general principle to
select, fix and stabilize the standard model parameters in the
way we do observe them. In fact, it should do much more: It
should choose the right gauge groups, the correct amount of
supersymmetry, the correct number of flavour families, it should
create an excess of matter over anti-matter, and so on. The
additional sector should fix the most relevant information about
the future universe at a very early stage,  similar as a DNA
string fixes the most important information of a human being
already when the first cells are formed.

We are still far away from such a theory. But some numerical
observations have recently been made \cite{physicad,book,prd} that
seem to give us a hint how this additional sector could look
like. Our aim here is to explain the relevant concepts in a
non-technical way.

In practice standard model parameters can be thought of as being
fixed by so-called moduli fields. A varying standard model
parameter (e.g. the fine structure constant) can be essentially
regarded as (a simple function of) such a moduli field. These
moduli fields evolve to minima of their potentials. So if we know
the correct moduli potentials describing the world around us, we
also know the correct standard model parameters. So what could be
a theory to construct these moduli potentials? In principle the
potentials should follow from the embedding theory (e.g. M theory
+ compactification + supersymmetry breaking), but little is known
in practice due to the enourmous complexity inherent in the above
theories. But as with any unknown theory we can be guided by
first trying to find an empirical theory that does the correct
thing, i.e. reproduces the observed value of the fine structure
constant and other fundamental constants, and then later try to
embed it into a greater context. The interesting thing is that
such an empirical theory is possible \cite{physicad,book,prd}:
There is a class of highly nonlinear chaotic dynamical system
that seem to reproduce the `correct' standard model parameters by
a simple selection mechanism, the minimization of vacuum energy.

Physically the above chaotic dynamics can be regarded as
describing rapidly fluctuating scalar fields associated with
vacuum fluctuations. These are vacuum fluctuations different from
those of QED or QCD, so what would be the physical embedding? The
most natural embedding is to associate the above chaotic vacuum
fluctuations with the currently observed dark energy in the
universe \cite{prd, spergel}. Since nobody really knows what dark
energy is there is a lot of freedom in the dark energy sector and
enough `space' to embed new things! The chaotic fields, living in
the dark energy sector, generate effective potentials for moduli
fields
--- just the same moduli fields that are responsible for the
fundamental constants of the standard model of electroweak and
strong interactions. The moduli fields then move to the minima of
the potentials generated by the chaotic fields, and fix the
fundamental constants of nature. The chaotic sector appears to
provide a possible answer to the question why we do observe
certain numerical values of standard model parameters (such as
the fine structure constant or the strong coupling at $W$-mass
scale) in nature, others not. It can be used to avoid anthropic
considerations for fundamental constants. Moreover, it generates
a small cosmological constant in a rather natural way
\cite{prd}.

These lecture notes are organized as follows: In section 2 we
provide some background information on moduli fields, variable
fine structure constants and related topics. In section 3 recall
the method of stochastic quantization introduced by Parisi and Wu
\cite{parisi, dam}. We then introduce the chaotic fields in section 4,
and show how they can generate potentials for moduli fields in
section 5. Finally, in section 6 we provide numerical evidence
that there are various local minima of the potentials that do
reproduce known standard model parameters with high precision. We
obtain excellent agreement with measured values seen in collider
experiments. We will also make a few predictions for unknown
parameters of the standard model, such as the Higgs mass, based
on the chaotic field dynamics and some additional symmetry
assumptions.

\section{Moduli fields and variable fine structure constant}

The action of the standard model is well known \cite{kane}, we
don't have to work that out here. The term of the action
involving the electromagnetic field tensor $F_{\mu\nu}$ is given
by
\begin{equation}
S_F= \int d^4 x \sqrt{-g} \left( -\frac{1}{4} F_{\mu\nu}F^{\mu\nu}
\right) .
\end{equation}
In models with a variable fine structure constant $\alpha (t)$
\cite{gardner, damour, sandvik, bekenstein, olive} the above
action is modified as follows:
\begin{equation}
S_F= \int d^4 x \sqrt{-g} \left( -\frac{1}{4} B_F (\chi /M^*)
F_{\mu\nu}F^{\mu\nu} \right) . \label{modac}
\end{equation}
Here $\chi (t)$ is a homogeneous scalar field, henceforth called
the {\em moduli field} (a related field in string theory is the
dilaton field). $M^*$ is a mass of the order of magnitude of the
reduced Planck mass. $B_F$ is a function that is in principle
determined by the embedding theory (e.g.\ string theory or M
theory). In Bekenstein's model \cite{bekenstein},
\begin{equation}
B_F(\chi /M^*)=e^{-2(\chi -\bar{\chi})/M^*},
\end{equation}
but other choices are possible as well. Changes in the field
$\chi (t)$ imply changes in the fine structure constant $\alpha
(t)$. The relation is
\begin{equation}
\frac{\alpha (t)}{\bar{\alpha}} = \frac{1}{B_F (\chi(t)/M^*)}
\label{BF}
\end{equation}
where $B_F(\bar{\chi}/M^*)=1$. Here $\bar{\alpha}$ is the
stationary value of $\alpha (t)$, and $\bar{\chi}$ the stationary
value of $\chi (t)$.

The interpretation of the above generalized action (\ref{modac})
is quite obvious: Normally, the electromagnetic field tensor would
contribute with the term $-\frac{1}{4}F_{\mu\nu}F^{\mu\nu}$ in the
action. This relation is now only satisfied in the stationary
case. Otherwise the strength of this term is given by a prefactor
$B_F$ that depends on the value of the field $\chi (t)$ at a
given time $t$. The fine structure constant $\alpha (t)$ is a
simple function of the moduli field $\chi (t)$ according to
eq.~(\ref{BF}).

For small displacements of the field $\chi (t)$ from its
stationary value we may write
\begin{equation}
B_F(\chi(t)/M^*)\approx B_F(\bar{\chi}/M^*)+\frac{\chi
(t)-\bar{\chi}}{M^*}B_F'(\bar{\chi}/M^*),
\end{equation}
where $B_F'$ is the derivative of the function $B_F$. We thus
obtain in leading order from eq.~(\ref{BF})
\begin{equation}
\alpha (t) =\bar{\alpha} \left( 1- B_F'(\bar{\chi}/M^*) \frac{\chi
(t) -\bar{\chi}}{M^*} \right) . \label{1}
\end{equation}
In Bekenstein's model, $B_F'(\bar{\chi}/M^*)=-2$. Clearly, the
fine structure constant approaches its stationary value
$\bar{\alpha}$ if the moduli field $\chi$ approaches its
stationary value $\bar{\chi}$. Using $B_F'(\bar{\chi}/M^*)=-2$ we
may also write eq.~(\ref{1}) as
\begin{equation}
\frac{1}{2} M^* \frac{\alpha (t)-\bar{\alpha}}{\bar{\alpha}}
=\chi (t)- \bar{\chi} \label{same}
\end{equation}
which shows that the fine structure constant and the moduli field
are basically the same.

To introduce a dynamics for the moduli field $\chi$, we need to
know its potential $V(\chi )$. In a Robertson-Walker metric the
dynamics is then given by
\begin{equation}
\ddot{{\chi}} +3H \dot{\chi} +\frac{\partial V}{\partial
\chi}=-\xi_m \frac{\rho_m}{M^*},
\end{equation}
where $H$ is the Hubble parameter and $\xi_m$ is the coupling of
the field $\chi$ to matter (in particular dark matter). $\rho_m$
is the matter density of the universe at a given time. In typical
models studied in the literature \cite{gardner}, the coupling
$\xi_m$ is very small.

Whereas in general the potential $V$ is unknown, near to the
equilibrium point $\bar{\chi}$ we may expand it as
\begin{equation}
V(\chi )=\frac{1}{2} m^2 (\chi -\bar{\chi})^2 +const
\end{equation}
so that
\begin{equation}
\frac{\partial V}{\partial \chi} =m^2 (\chi -\bar{\chi}).
\end{equation}
In \cite{gardner} $m$ is chosen as an extremely small mass
parameter, of the order of the current value $H_0$ of the Hubble
parameter:
\begin{equation}
m \sim H_0
\end{equation}
That is to say, one considers an ultralight scalar field $\chi$
with a mass of the order $10^{-33}$ eV. There is motivation for
the existence of such ultralight fields from extended supergravity
theories \cite{kallosh}. In terms of the fine structure constant
$\alpha (t)$, one obtains near the stationary point
\begin{equation}
V(\chi) = \frac{1}{2} m^2 {M^*}^2 \frac{1}{(B_F'
(\bar{\chi}/M^*))^2} \frac{(\alpha
(t)-\bar{\alpha})^2}{\bar{\alpha}^2} +const
\end{equation}
By construction, the energy density $m^2{M^*}^2$ associated with
this potential is of the same order of magnitude as the dark
energy density of the universe at the present time.

Of course, in a similar way one can introduce further moduli
fields corresponding to other standard model coupling constants,
masses and mixing parameters as well. The fine structure constant
was just one example. In fact, for each relevant standard model
parameter there should be a corresponding moduli field. So we
need about 20 such fields.

\section{Stochastic quantization}

Let us now proceed to a 2nd-quantized theory. An elegant method to
do 2nd quantization is via the so-called stochastic quantization
method. In the Parisi-Wu approach of stochastic quantization one
considers a stochastic differential equation evolving in a
fictitious time variable $s$, the drift term being given by the
classical field equation \cite{parisi, dam}. Quantum mechanical
expectations correspond to expectations with respect to the
generated stochastic processes in the limit $s\to \infty$. The
fictitious time $s$ is different from the physical time $t$, it
is just a helpful fifth coordinate to do 2nd quantization.
Neglecting spatial gradients the field under consideration is a
function of physical time $t$ and fictitious time $s$. The
stochastically quantized equation of motion of a homogeneous
scalar field $\varphi$ in Robertson-Walker metric is
\begin{equation}
\frac{\partial}{\partial s}\varphi =\ddot{\varphi}
+3H\dot{\varphi} +V'(\varphi) +L(s,t), \label{sto}
\end{equation}
where $H$ is the Hubble parameter, $V$ is the potential under
consideration and $L(s,t)$ is Gaussian white noise,
$\delta$-correlated both in $s$ and $t$. For e.g.\ a numerical
simulation we may discretize eq.~(\ref{sto}) using
\begin{eqnarray}
s &=& n\tau \\ t &=& i \delta ,
\end{eqnarray}
where $n$ and $i$ are integers and $\tau$ is a fictitious time
lattice constant, $\delta$ is a physical time lattice constant.
The continuum limit requires $\tau \to 0$, $\delta \to 0$, but
since quantum field theory is expected to break down at the
Planck scale, it can make physical sense to keep a small lattice
constant of that size as an effective cutoff:
\begin{equation}
\tau \sim \frac{1}{m_{Pl}^2}, \delta \sim \frac{1}{m_{Pl}}.
\end{equation}
Eq.~(\ref{sto}) yields the discrete dynamics
\begin{equation}
\frac{\varphi_{n+1}^i-\varphi_n^i}{\tau} = \frac{1}{\delta^2}
(\varphi_n^{i+1}-2\varphi_n^i+\varphi_n^{i-1}) +3\frac{H}{\delta}
(\varphi_n^i-\varphi_n^{i-1}) +V'(\varphi_n^i) + noise .
\end{equation}
This can be written as the following recurrence relation for the
field $\varphi_n^i$
\begin{equation}
\varphi_{n+1}^i= (1-a)\left\{ \varphi_n^i
+\frac{\tau}{1-a}
V'(\varphi_n^i)\right\}+3\frac{H\tau}{\delta} (\varphi_n^i-
\varphi_n^{i-1})+\frac{a}{2}(\varphi_n^{i+1}+\varphi_n^{i-1})
+ \tau\cdot noise,
\end{equation}
where a dimensionless coupling constant $a$ is introduced as
\begin{equation}
a:=\frac{2\tau}{\delta^2}.
\end{equation}
This coupling $a$ is a free parameter of our stochastically
quantized theory. A priori it can take on any value, reminescent
of a moduli field.

We may also introduce a dimensionless field variable $\Phi_n^i$
by writing $\varphi_n^i=\Phi_n^i p_{max}$, where $p_{max}$ is
some (so far) arbitrary energy scale. The above scalar field
dynamics is equivalent to a spatially extended dynamical system
(a coupled map lattice) of the form
\begin{equation}
\Phi_{n+1}^i=(1-a)T(\Phi_n^i)+\frac{3}{2}H\delta a
(\Phi_n^i-\Phi_n^{i-1})+\frac{a}{2}(\Phi_n^{i+1}+\Phi_n^{i-1}) +
\tau\cdot noise, \label{dyni}
\end{equation}
where the local map $T$ is given by
\begin{equation}
T(\Phi )=\Phi
+\frac{\tau}{p_{max}(1-a)}V'(p_{max}\Phi).\label{map}
\end{equation}
Here the prime means
\begin{equation}
'=\frac{\partial}{\partial
\varphi}=\frac{1}{p_{max}}\frac{\partial}{\partial \Phi}.
\end{equation}
A symmetric dynamics of the form
\begin{equation}
\Phi_{n+1}^i=(1-a)T(\Phi_n^i)+\frac{a}{2}(\Phi_n^{i+1}+\Phi_n^{i-1})
+\tau \cdot noise \label{sym}
\end{equation}
is obtained if $H\delta << 1$, equivalent to
\begin{equation}
\delta << H^{-1}.
\end{equation}
This approximation is valid if the universe is much older than
the physical time lattice constant $\delta$. In this case the
term proportional to $H$ in eq.~(\ref{dyni}) can be neglected. The
local map $T$ depends on the potential $V$ under consideration.
Since we restrict ourselves to real scalar fields $\varphi$, $T$
is a 1-dimensional map.

Let us summarize the main result of this section: Iterating a
recurrence relation of the form (\ref{sym}) is {\em equivalent} to
considering a stochastically quantized scalar field. The relation
between the map $T$ and the potential $V$ is
\begin{equation}
V'(\varphi)=\frac{1-a}{\tau} \left\{ -\varphi +p_{max}
T\left(\frac{\varphi}{p_{max}}\right) \right\}.
\end{equation}
Integration yields
\begin{equation}
V(\varphi) =\frac{1-a}{\tau} \left\{ -\frac{1}{2} \varphi^2 +
p_{max}\int d\varphi \,T\left(\frac{\varphi}{p_{max}}\right)
\right\} + const ,\label{pot}
\end{equation}
which in terms of the dimensionless field $\Phi$ this can be written as
\begin{equation}
V(\varphi)=\frac{1-a}{\tau} p_{max}^2 \left\{ -\frac{1}{2}
\Phi^2+\int d\Phi T(\Phi) \right\} +const.
\end{equation}

\section{Introducing chaotic fields}

We now come to the crucial point, namely what type of dynamics is
generated by (\ref{sym}) for various types of scalar field
theories. Take the example of an ultralight moduli field $\chi$.
For these types of fields $p_{max} \sim M^* \sim m_{Pl}$ is
large, whereas $V'(\chi) \sim m^2(\chi -\bar{\chi})$ is extremely
small. This means the mapping $T$ in eq.~(\ref{map}) is extremely
close to an identity and the field moves very slowly. Effectively
this means that our discrete dynamics (\ref{sym}) approximates
very well a smooth continuum evolution of the field $\chi$. The
field $\chi$ smoothly approaches a minimum of the potential and
stays there, apart from some small fluctuations induced by the
noise term.

Once again, for a moduli field $\chi$ the energy scale associated
with the field variable is very large, of the order of the Planck
mass $M^*\sim m_{Pl}$, whereas the potential
$V$ contains an extremely small mass parameter $m$, of the order
of the present Hubble constant $H_0$. We could ask whether for
symmmetry reasons maybe another scalar field $\varphi$ exists
that has just the opposite properties, i.e. the energy scale of
that field variable $\varphi$ is extremely small (of the order
$H_0$) but its potential $V$ contains a mass term of order $M^*$?

Indeed, such a field has been studied in \cite{prd}. Due to the
fact that the forcing is now very strong, this field exhibits
strongly chaotic behaviour if it is stochastically quantized. This
can immediately be seen from eq.~(\ref{map}): With $p_{max}$ being
small, the map $T$ is now far away from the identity. Chaotic
behaviour is possible. The small noise term in eq.~(\ref{sym})
can actually be neglected in the chaotic case.

We may go one step further in our symmetry considerations. The
moduli field is as slow and regular as a field can be. What would
be the other extreme? What would be a field that is as rapidly
fluctuating and
irregular as possible? In other words, which scalar field
dynamics would create the strongest possible chaotic behaviour?

The above question has been solved in \cite{nonli,hilgers} and is
well understood from a nonlinear dynamics point of view. One knows
that the maps with strongest chaotic properties (being smooth and
deterministic at the same time) are those conjugated to a
Bernoulli shift \cite{BS} (a shift of integer symbols in suitable
coordinates).  Among those, certain deterministic maps are even
more `random' than others, in the sense of having least
higher-order correlations: These are the so-called Tchebyscheff
maps $T_N$ of $N$-th order, defined as
\begin{eqnarray}
T_2 (\Phi)&=&2\Phi^2-1 \\
T_3 (\Phi) &=&4\Phi^3-3\Phi \\
... &=& ... \\
T_N (\Phi) &=& \cos (N \arccos \Phi)
\end{eqnarray}
They can arise out of the above dynamics (\ref{sym}) for suitable
potentials $V$.

The most important scalar field potential in particle physics is
of course a double-well potential. So let us consider the
distinguished example of a $\varphi^4$-theory generating
strongest possible chaotic behaviour. Take the potential
\begin{equation}
V_{-3}(\varphi)=\frac{1-a}{\tau}\left\{
\varphi^2-\frac{1}{p_{max}^2} \varphi^4\right\}+const, \label{16}
\end{equation}
or, in terms of the dimensionless field $\Phi$,
\begin{equation}
V_{-3}(\varphi)=\frac{1-a}{\tau} p_{max}^2 ( \Phi^2 -\Phi^4)
+ const. \label{17}
\end{equation}
Applying our formalism of the previous section, we end up with the
following local map:
\begin{equation}
\Phi_{n+1}=T_{-3}(\Phi_n)=-4\Phi_n^3+3\Phi_n
\end{equation}
$T_{-3}$ is the (negative)
third-order Tchebyscheff map.
It is conjugated to a
Bernoulli shift of 3 symbols, and generates the strongest possible stochastic
behaviour possible for a smooth low-dimensional deterministic
dynamical system.

Apparently, starting from the potential (\ref{16}) we obtain by
second quantization a field $\varphi$ that rapidly fluctuates in
fictitious time on some finite interval, provided that initially
$\varphi_0\in [-p_{max},p_{max}]$. The small noise term in
eq.~(\ref{sym}) can be neglected as compared to the deterministic
chaotic fluctuations of the field. We physically interprete these
rapid changes of the field $\varphi$ as representing vacuum
fluctuations. Of course these are vacuum fluctuations different
from those of QED or QCD. Since the expectation of the vacuum
energy associated with the chaotic field is $\langle
V_{-3}(\varphi) \rangle \sim p_{max}^2/\tau \sim H_0^2 m_{Pl}^2$,
such a chaotic field yields the correct order of magnitude of
vacuum energy density in order to account for a small
cosmological constant. Hence we assume that the chaotic fields
underly dark energy.

The above example of a chaotic $\varphi^4$-theory can be
generalized. There are various discrete degrees of freedom to
introduce a deterministic chaotic field dynamics that is as
random as possible. Consider a 1-dimensional lattice, the lattice
sites are labelled by integers $i$. At each lattice site $i$ we
have a dimensionless field variable $\Phi_n^i$ which evolves in
discrete time $n$. In the uncoupled case, the dynamics is given by
\begin{equation}
\Phi_{n+1}^i= \pm T_N(\Phi_n^i),
\end{equation}
where $\pm T_N$ is either the positive of the negative
Tchebyscheff map. The initial value $\Phi_0$ is chosen on the
interval $[-1,1]$, the iterates $\Phi_n^i$ then stay in this
interval, but evolve in a deterministic chaotic way.
The dynamics is conjugated to a Bernoulli shift of $N$ symbols,
which means that in suitable coordinates the iteration process is
like shifting symbols in a random symbol sequence. There are
further discrete degrees of freedom to do the coupling to the
nearest neighbours. Instead of coupling to the variables
$\Phi_n^{i\pm 1}$ at the neighboured sites as in eq.~(\ref{sym}),
we could also use the updated variables $\pm T_N(\Phi_n^{i\pm
1})$. Since $n$ describes fictitious time, it's really not clear
which choice is the correct one. For sure both degrees of freedom
exist. All these discrete degrees of freedom can be written in a
compact form as follows:
\begin{equation}
\Phi_{n+1}^i=(1-a)T_N(\Phi_n^i) + s \frac{a}{2} (T_N^b
(\Phi_n^{i-1}) +T_N^b (\Phi_n^{i+1})), \label{dyn}
\end{equation}
where $s$ is a sign variable taking on the values $\pm1$. The
choice $s =+ 1$ is called `diffusive coupling', but for symmetry
reasons it also makes sense sense to study the choice $s=-1$,
which we call `anti-diffusive coupling'. The integer $b$
distinguishes between the forward and backward coupling form,
$b=1$ corresponds to forward coupling ($T_N^1(\Phi):=T_N(\Phi))$,
$b=0$ to backward coupling ($T_N^0(\Phi):=\Phi$). We consider
random initial conditions, periodic boundary conditions and large
lattices of size $i_{max}$.

One can easily check that for odd $N$ the choice of $s$ is
irrelevant (since odd Tchebyscheff maps satisfy $T_N(-\Phi
)=-T_N(\Phi )$), whereas for even $N$ the sign of $s$ is relevant
and a different dynamics arises. Hence, restricting ourselves to
$N=2$ and $N=3$, in total 6 relevant chaotic scalar field theories
arise, characterized by
$(N,b,s)=(2,1,+1),(2,0,+1),(2,1,-1),(2,0,-1)$ and
$(N,b)=(3,1),(3,0)$. For easier notation, in the following we will
label these chaotic field theories as $2A,2B,2A^-,2B^-,3A,3B$,
respectively.

The important thing to remember from this section is that
eq.~(\ref{dyn}) just describes a degenerated stochastically
quantized scalar field with strongest possible chaotic properties.
For somebody from the nonlinear dynamics and complexity community,
the dynamics (\ref{dyn}) represents the most natural thing in the
world: It's just a coupled map lattice, a standard example of a
spatio-temporal dynamical system \cite{kaneko}. For somebody from
the elementary particle physics or string theory community,
however, eq.~(\ref{dyn}) may look somewhat unusual and unfamiliar
at first sight. But it's sometimes worth to learn new things!

The dynamics (\ref{dyn}) is a dissipative deterministic dynamical
system. It exhibits chaotic behaviour and produces information,
measured by a positive KS entropy \cite{BS}. Dissipative
deterministic systems as a model of quantum gravity have also
been suggested in \cite{thooft}.

\section{Generating potentials for moduli fields}

Clearly, if the coupling $a$ in the above chaotic field dynamics
(\ref{dyn}) is chosen as $a=0$, then there are no correlations
between neighboured lattice sites:
\begin{equation}
\langle \Phi_n^i \Phi_n^{i+1} \rangle =0 \label{co}
\end{equation}
Here the notation $\langle \cdots \rangle$ denotes the expectation
value, which can be numerically calculated by iterating the maps
for random initial conditions and doing a time average. Note that
the index $i$ denotes physical time (in units of the lattice
constant $\delta$). If we physically interpret the chaotic
fluctuations of the field $\Phi_n^i$ as some sort of vacuum
fluctuations underlying the dark energy of the universe (see
\cite{prd} for more details), then the above condition (\ref{co})
looks physically very reasonable: We want subsequent vacuum
fluctuations to be uncorrelated in physical time, because
otherwise they wouldn't really describe spontaneous fluctuations.

We may, however, insist on an interacting theory, i.e.\ $a \not=
0$. Are there still some distinguished couplings $a^*\not= 0$
where we can keep the above condition of a vanishing correlation
function in physical time? Of course such a state would again be
distinguished as being as random as possible, where this concept
is now extended from fictitious time to physical time. There are
also a couple of other reasons why one wants a vanishing
correlation function, see \cite{book} for more details.

In fact, numerical investigations show that the above states with
vanishing correlation exist and distinguish certain coupling
constants $a^*$. These are numerically observed to coincide with
known standard model coupling strengths (see next section).
 In general one can show \cite{physicad, book}
that the quantity
\begin{equation}
W(a)=\frac{1}{2} \langle \Phi_n^i \Phi_n^{i+1} \rangle
\end{equation}
 can be physically interpreted as the interaction energy of
the chaotic field theory under consideration (in suitable units).
Hence states of strongest random properties, described by a
vanishing correlation function, have vanishing interaction energy.

Besides the interaction energy,
there is also another relevant vacuum energy associated with
the chaotic fields. This is the self energy $V(a)$, given by
\begin{equation}
V^{(2)}(a)= \langle \Phi \rangle -\frac{2}{3}\langle \Phi^3 \rangle \;\;\;\; (N=2), \label{39}
\end{equation}
respectively
\begin{equation}
V^{(3)}(a)= \frac{3}{2} \langle \Phi^2 \rangle - \langle \Phi^4 \rangle \;\;\;\; (N=3). \label{40}
\end{equation}
For a derivation, see \cite{book}. The self energy describes the
vacuum energy associated with the potentials that generate the
chaotic dynamics in fictitious time (the additive constant is
fixed by some symmetry considerations).

In the following, we will use both the interaction energy
and the self energy to generate suitable potentials for moduli fields.
Recall that classically the moduli field $\chi$ obeys


\begin{equation}
\ddot{\chi} +3H\dot{\chi} +
  V'(\chi )=0, \label{fieldeq}
\end{equation}
where we neglected possible interactions with dark matter.
In the vicinity of the stationary state $\bar{\chi}$, one has $V'(\chi)=m^2 (\chi -\bar{\chi})$ and
the moduli field is essentially the same as the standard
model coupling constant $\alpha$. According to eq.~(\ref{same}) we have
\begin{equation}
\chi (t)= \frac{M^*}{2\bar{\alpha}}\alpha (t)+\bar{\chi}-\frac{1}{2} M^*. \label{same2}
\end{equation}
Putting eq.~(\ref{same2}) into (\ref{fieldeq}), we
obtain an equation for $\alpha$,
\begin{equation}
\ddot \alpha +3H \dot{\alpha} + V'(\alpha )=0, \label{harm}
\end{equation}
where locally the potential is given by
\begin{equation}
V(\alpha)=\frac{1}{2}m^2 (\alpha - \bar{\alpha})^2.
\end{equation}
Eq.~(\ref{harm}) is just the equation of a damped harmonic oscillator,
provided $\alpha$ is in
the vicinity of $\bar{\alpha}$.

The crucial point is now that we assume that {\em the same}
moduli fields are responsible for the coupling constants  of the
standard model of electroweak and strong interactions and  those
of the dark sector described by the chaotic fields. This means
$a=\alpha$. The chaotic field wants to find a state of strongest
possible random properties described by a vanishing correlation
function $\langle \Phi_n^i \Phi_n^{i+1} \rangle$. Hence the
moduli field $\chi$ given by eq.~(\ref{same2}) with $\alpha =a$
needs to adjust its value. This can be achieved by choosing in
eq.~(\ref{harm}) the formal potential\footnote{Our sign
conventions relate to moduli potentials generated by {\em
positive} Tchebyscheff maps.}
\begin{equation}
V(\alpha)= - m^2 \int_0^\alpha d\alpha' \langle \Phi_n^i
\Phi_n^{i+1} \rangle . \label{entro}
\end{equation}
By construction, local minima of this potential are stable
stationary points corresponding to stable (attracting) zeros of
the forcing $V'(\alpha)= - m^2 \langle \Phi_n^{i}\Phi_n^{i+1}
\rangle$ (see Fig.~1). If the damping $H$ in eq.~(\ref{harm}) is
large as compared to $m$, and if the initial displacement is not
too large, then the stationary state $\bar{\alpha}$ is rapidly
approached.

\begin{figure}
\epsfig{file=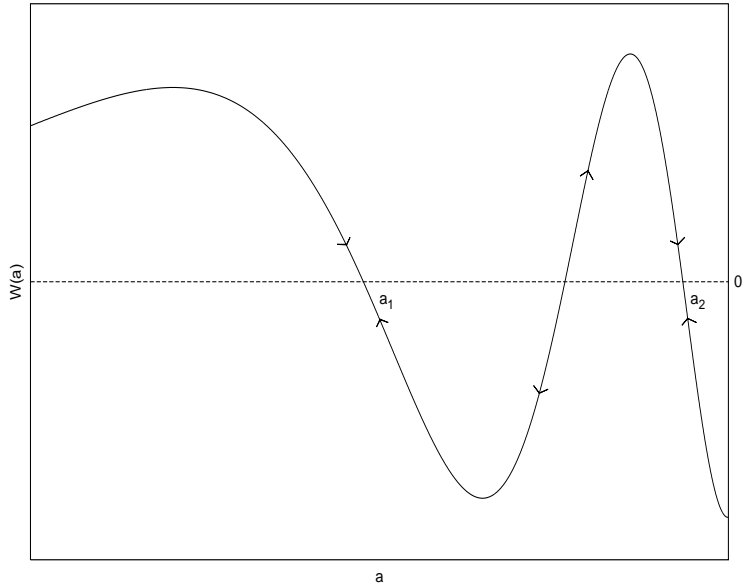, width= 10cm, height =8cm}
\caption{Basic idea underlying this paper. A priori all kinds of
standard model couplings $\alpha = a$ (values of moduli fields)
are possible. These are then driven into stable zeros of the
interaction energy $W(a)$ of the chaotic fields. In the above
schematic picture, the stable zeros are denoted by $a_1$ and
$a_2$.}
\end{figure}

For other types of moduli fields, we may choose in eq.~(\ref{harm})
the self energy potential $V(\alpha) =m^2 V^{(N)}(\alpha)$
as given in eq.~(\ref{39}) or (\ref{40}). These types of moduli fields then evolve to local minima of
the self energy.


It is well known that standard model interaction strengths
actually depend on the relevant energy scale $E$. We have the
running electroweak and strong coupling constants. For example,
the fine structure constant $\alpha_{el}(E)$ slightly increases
with $E$, and the strong coupling $\alpha_s$ rapidly decreases
with $E$.

What should we now take for the energy (or temperature) $E$ of
the moduli fields near to their stationary state?
In other words, if a standard model coupling $\bar{\alpha}$ is fixed as
a minimum of the potential $V(\alpha)$,
at which energy scale $E$ is this standard model coupling fixed?
{\em A priori} this is unknown. However, there is extensive
numerical evidence \cite{physicad, book} that the distinguished
couplings $\bar{\alpha}=a^*$ which correspond to minima of the
potentials numerically coincide with running standard model
coupling constants $\alpha (E)$ with an energy $E$ being given by
\begin{equation}
E= \frac{1}{2} N (m_{B}+m_{f_1}+m_{\bar{f_2}}). \label{katharina}
\end{equation}
Here $N$ is the index of the Tchebyscheff map considered, and
$m_{B}, m_{f_1},m_{\bar{f_2}}$ denote the masses of a boson $B$
and a fermion $f_1$ and anti-fermion $\bar{f_2}$ ---not just some
exotic bosons and fermions but precisely those that we know from
the standard model. One typically observes particle combinations
$B,f_1,\bar{f_2}$ that describe possible interaction states in
the standard model, for example a decay of the form $B\to
f_1+\bar{f_2}$ or a reaction $f_1+\bar{f_2} \to B$. Detailed
evidence will be given in the next section. Formula
(\ref{katharina}) formally reminds us of the energy levels
$E_N=\frac{N}{2} \hbar \omega$ of a quantum mechanical harmonic
oscillator, with low-energy levels ($N=2,3$) given by the masses
of the standard model particles.

\section{Numerical results}

We now present our numerical results (much more details can be
found in \cite{physicad, book}). In particular, we will show that
stable zeros of the correlation functions (states of strongest
random properties for chaotic fields, which are stationary states
for moduli fields) reproduce known standard model coupling
constants and masses with high precision. Our results are
obtained by long-term iteration of the chaotic dynamics. One
numerically calculates, for a given coupling $a$, the interaction
energy as the time average
\begin{equation}
W(a)= \frac{1}{2} \langle \Phi_n^i \Phi_n^{i+1} \rangle
=\frac{1}{2} \lim_{n\to \infty} \lim_{J\to \infty} \frac{1}{MJ}
\sum_{n=1}^M \sum_{i=1}^J \Phi_n^i \Phi_n^{i+1},
\end{equation}
where the $\Phi_n^i$ evolve according to (\ref{dyn}) with random
initial conditions. In practice, we used a finite lattice of size
$J=10000$ with periodic boundary conditions, and iteration numbers
corresponding to several weeks of CPU time. Everybody is welcome
to reproduce and verify the numerical results described
below---the recurrence relation (\ref{dyn}) can be very easily
installed on any computer. What one observes is that the stable
zeros of $W(a)$ coincide with known standard model interaction
strengths, thus indicating the physical relevance of the theory
presented presented in sections 1--5. The chaotic fields appear
to determine the moduli potentials in precisely the way we need
them to be for a physically realistic vacuum. They can `help' the
universe to find the `right' vacuum out of an incredibly large
number of choices. Anthropic considerations can be avoided in
this context.

\subsection{
The 3A dynamics---electric interaction strengths of electrons and
$d$-quarks} Fig.~2 shows the interaction energy $W(a)=\frac{1}{2}
\langle \Phi_n^i \Phi_n^{i+1} \rangle$ of the chaotic $3A$
dynamics in the small-coupling region.  We observe two stable
zeros of the interaction energy in the low-coupling region:
\begin{figure}
\epsfig{file=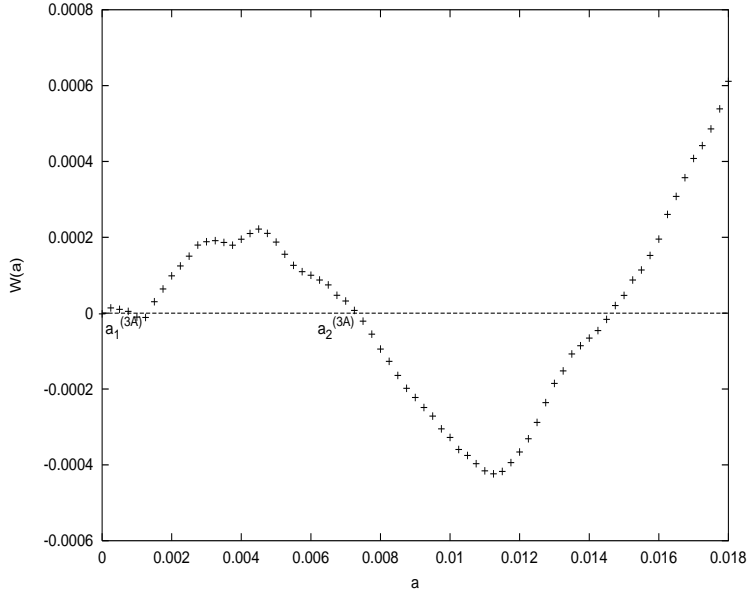, width=10cm, height=8cm}
\caption{Interaction energy of the 3A dynamics in the
small-coupling region.}
\end{figure}
\begin{eqnarray*}
a_1^{(3A)}&=&0.0008164(8)     \\ a_2^{(3A)}&=&0.0073038(17)
\end{eqnarray*}
Previously these stable zeros were denoted by $a^*$ or
$\bar{\alpha}$. A stable zero satisfies $V'(a^*)=0$ and
$V''(a^*)>0$, with $V$ given by eq.~(\ref{entro}).

Remarkably, the zero $a_2^{(3A)}$ appears to numerically coincide with the
fine structure constant $\alpha_{el} \approx 1/137$. To construct
a suitable physical interpretation in the sense of
eq.~(\ref{katharina}), let us choose  $B=\gamma$,
$f_1=e^-,\bar{f}_2=e^+$. The relevant energy scale underlying
this moduli field is then given by $E=(3/2)(m_\gamma+2m_e)=3m_e$,
according to eq.~(\ref{katharina}). Hence our standard model
interpretation of the stationary moduli state described by
$a_2^{(3A)}$ suggests the numerical identity
\begin{equation}
a_2^{(3A)}=\alpha_{el}(3m_e).
\end{equation}
For a precise numerical comparison
let us estimate the running electromagnetic coupling at this energy scale.
We may use the 1st-order QED formula
\begin{equation}
\alpha_{el} (E) = \alpha_{el} (0) \left\{ 1+ \frac{2\alpha_{el}
(0)}{\pi} \sum_i f_i \int_0^1dx\; x(1-x)  \log \left( 1+
\frac{E^2}{m_i^2}x(1-x)\right) \right\}. \label{runael}
\end{equation}
The sum is over all charged elementary particles, $m_i$ denotes
their (free) masses, and $f_i$ are charge factors given by 1 for
$e,\mu ,\tau$-leptons, $\frac{4}{3}$ for $u,c,t$-quarks and
$\frac{1}{3}$ for $d,s,b$-quarks. Using this formula, we get
$\alpha_{el}(3m_e) =0.007303$, to be compared with
$a_2^{(3A)}=0.0073038(17)$. There is excellent agreement.
Inverting the argument, the above zero of the interaction energy
of the chaotic field can be used to predict
the numerical value of the fine structure constant from first principles.

Next, we notice that the other zero $a_1^{(3A)}$ has approximately the value
$\frac{1}{9} \alpha_{el}$. This could mean that the chaotic 3A
field also has a mode that provides evidence for electrically
interacting $d$-quarks. Our interpretation is
\begin{equation}
a_1^{(3A)} =\alpha_{el}^d (3 m_d)=\frac{1}{9} \alpha_{el} (3 m_d),
\end{equation}
where $\alpha_{el}^d=\frac{1}{9}\alpha_{el}$ denotes the
electromagnetic interaction strength of $d$-quarks. In the
harmonic oscillator interpretation (\ref{katharina}), we may
choose $B=\gamma$, $f_1=d,f_2=\bar{d}$. Formula (\ref{runael}),
as an estimate, yields for $m_d\approx  9$ MeV the value
$\alpha_{el}(3m_d)= 0.007349$, which coincides very well with
$9a_1^{(3A)}=0.007348(7)$. The value $9a_1^{(3A)}$ actually
translates to the energy scale $E_d=(26.0 \pm 6.4)$ MeV. This
yields $m_d = \frac{1}{3}E_d=(8.7 \pm 2.1)$ MeV, which coincides
with estimates of the $\overline{MS}$ current quark mass of the
$d$ quark at the proton mass renormalization scale.


\subsection{The 3B dynamics ---weak interaction strengths
of neutrinos and $u$-quarks}

The interaction energy $W(a)$ of the 3B field is plotted in
Fig.~3. In the low-coupling region $a\in[0,0.018]$ we observe the
following stable zeros of $W(a)$:
\begin{figure}
\epsfig{file=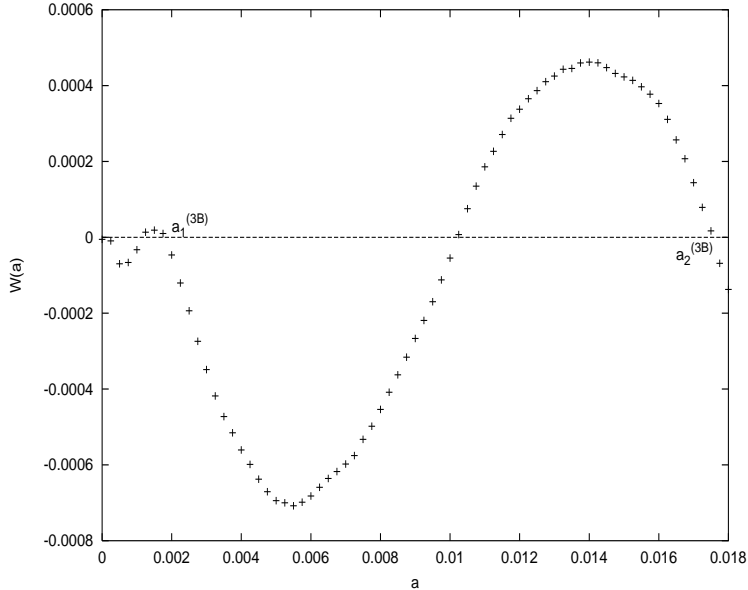, width=10cm, height=8cm}
\caption{Interaction energy of the 3B dynamics in the
low-coupling region.}
\end{figure}
\begin{eqnarray*}
a_1^{(3B)} &=& 0.0018012(4) \\ a_2^{(3B)} &=& 0.017550(1)
\end{eqnarray*}
If our approach is consistent, we should be able to find an
interpretation of $a_1^{(3B)}$ and $a_2^{(3B)}$ in terms of
moduli fields fixing the standard model coupling strengths of
$u$-quarks and neutrinos.

Let us start with $a_2^{(3B)}$. For left-handed neutrinos, the weak
coupling due to the exchange of $Z^0$-bosons is given by
\begin{equation}
\alpha_{weak}^{\nu_L}=\alpha_{el} \frac{1}{4 \sin^2 \theta_W \cos^2 \theta_W}.
\end{equation}
Here $\theta_W$ is the weak mixing angle. In the following we will
treat $\sin^2 \theta_W $ as an effective constant, and regard
$\alpha_{el}$ as the running electromagnetic coupling. Other
renormalization schemes are also possible, but yield only minor
numerical differences. Experimentally, the effective weak mixing
angle is measured as $\sin^2 \theta_W=\bar{s}_l^2\approx 0.2315$
\cite{pada}. Assuming that in addition to the left-handed
neutrino interacting weakly there is an electron interacting
electrically, the two interaction processes can add up
independently if the electron is right-handed, since right-handed
electrons cannot interact with left-handed neutrinos. Hence a
possible standard model interpretation of the zero $a_2^{(3B)}$
would be
\begin{equation}
a_2^{(3B)}=\alpha_{el}(3 m_e) +\alpha_{weak}^{\nu_L} (3 m_{\nu_e} )
=a_2^{(3A)}+ \alpha_{el} (3m_{\nu_e}) \frac{1}{4\sin^2 \theta_W \cos^2
\theta_W} \label{61}
\end{equation}
In the harmonic oscillator interpretation of
eq.~(\ref{katharina}), we choose $B$ massless, $f_1=\nu_L$,
$\bar{f}_2=\bar{\nu_L}$ in addition to the process already
described by $a_2^{(3A)}$. Putting in the experimentally measured
value of $\sin^2\theta_W=0.2315$, we obtain for the right-hand
side of eq.~(\ref{61}) the value 0.01756, which coincides to 4
digits with the observed stationary value of the moduli field
$a_2^{(3B)}=0.01755$.

Next, let us interpret $a_1^{(3B)}$. In analogy to the joint
appearance of $\nu$ and $e$, we should also expect to find evidence for a
weakly interacting $u$-quark, together with a $d$-quark
interacting electrically. Clearly, the $u$-quark could also
interact electrically, but for symmetry reasons we expect the pair
$(u,d)$ to interact in a similar way as $(\nu ,e)$. A right-handed
$u$-quark interacts weakly with the coupling
\begin{equation}
\alpha_{weak}^{u_R} = \frac{4}{9} \alpha_{el} \frac{\sin^2 \theta_W}{
\cos^2 \theta_W}.
\end{equation}
Adding up the electrical interaction strength of a $d$-quark, a
natural interpretation, quite similar to that of the zero
$a_2^{(3B)}$, is
\begin{equation}
a_1^{(3B)}=\alpha_{el}^{d} (3 m_d) + \alpha_{weak}^{u_R}(3 m_u)
 = a_1^{(3A)}+ \frac{4}{9} \alpha_{el} (3 m_u)
\frac{\sin^2 \theta_W}{\cos^2 \theta_W}  \label{america}
\end{equation}
The harmonic oscillator interpretation of this moduli state is $B$
massless, $f_1=u_R$, $f_2=\bar{u_R}$ in addition to the process
underlying $a_1^{(3A)}$. Numerically, taking $\sin^2
\theta_W=0.2315$ and evaluating the running $\alpha_{el}$ using
$m_u \approx 5$ MeV, we obtain for the right-hand side of
eq.~(\ref{america}) 0.001800, which should be compared with
$a_2^{(3B)}=0.001801$. Again we have perfect agreement within the
first 4 digits. It is remarkable that the same universal effective
value $\sin^2 \theta_W =0.2315$ can be used consistently for both
leptons (couplings $a_2^{(3A)}$, $a_2^{(3B)})$ and quarks
(couplings $a_1^{(3A)}$, $a_1^{(3B)}$).
Inverting the formulas, the observed stationary values of the
moduli fields can be used to predict that the weak mixing angle
is $\bar{s}_l^2\approx 0.2315$.

Note that generally the backward coupling form of the $N=3$
chaotic fields seems to describe
a spinless state (formed by $e_R$ and $\nu_L$, respectively $d_L$
and $u_R$), whereas the forward coupling form just describes one
particle species with non-zero spin ($e$ or $d$). A similar statement will
turn out to hold for
the $N=2$ theories, replacing fermions by bosons.


\subsection{The 2A dynamics
---strong interaction strength at the $W$-mass scale}
If electroweak coupling strengths are fixed by suitable moduli
potentials generated by chaotic fields, then something similar
should also be the case for the strong coupling strength
$\alpha_s$. Let us now look at chaotic fields with $N=2$. Fig.~4
shows the interaction energy $W(a)$ of the 2A dynamics. Only one
stable zero is observed:
\begin{figure}
\epsfig{file=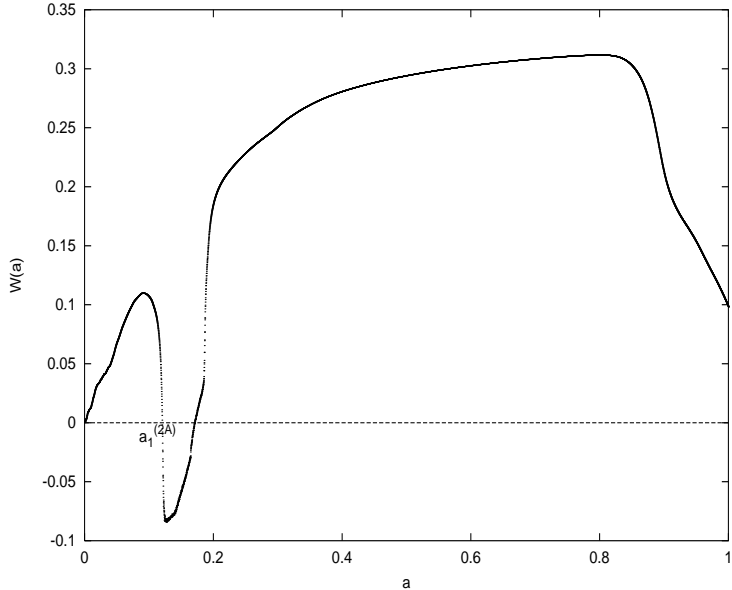, width=10cm, height=8cm}
\caption{Interaction energy of the 2A dynamics.}
\end{figure}
\begin{equation}
a_1^{(2A)} = 0.120093(3) \label{strong}
\end{equation}
We notice that it numerically seems to coincide with the strong
coupling constant $\alpha_s$ at the $W$- or $Z$ mass scale, which
is experimentally measured as $\alpha_s (m_Z)\approx 0.118$.

For symmetry reasons, it seems plausible that if the $N=3$
dynamics fixes the electroweak couplings at the smallest
fermionic mass scales, then the $N=2$ dynamics could fix the
strong couplings at the smallest bosonic mass scales. The
lightest massive gauge boson is indeed the $W^\pm$. Hence our
physical interpretation associated with $a_1^{(2A)}$ would be
$B=W^\pm$, $f_1=u$, $\bar{f_2}=\bar{d}$ (respectively $f_1=d,
\bar{f_2}=\bar{u}$), and since $N=2$ formula (\ref{katharina})
implies
\begin{equation}
a_1^{(2A)}=\alpha_s(m_W+m_u+m_d) \approx \alpha_s (m_W).
\label{Franziska}
\end{equation}
Since the $W$-mass is known with high precision, eq.~(\ref{Franziska})
and (\ref{strong}) yield
quite a precise prediction for the strong coupling $\alpha_s$. We can
evolve it with high precision to arbitrary energy scales,
using the well known perturbative formulas from QCD, obtaining
\begin{equation}
\alpha_s (m_{Z^0})=0.117804(12).
\end{equation}
This prediction of $\alpha_s$ from the zero of the chaotic 2A
dynamics is clearly consistent with the experimentally measured
value 0.118 and in fact much more precise than current
experiments can verify.

\subsection{The 2B dynamics---the lightest scalar glueball}
The interaction energy of the 2B dynamics is shown in Fig.~5.
$W(a)$ has only one non-trivial zero
\begin{figure}
\epsfig{file=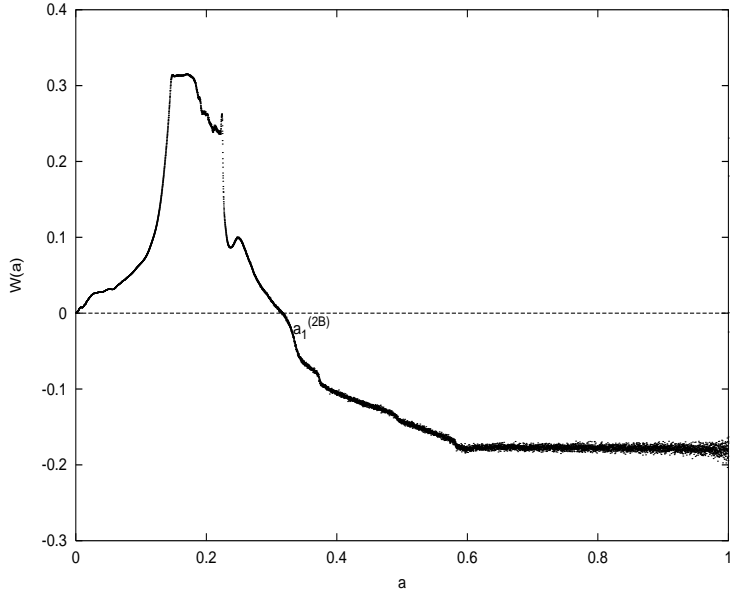, width=10cm, height=8cm}
\caption{Interaction energy of the 2B dynamics.}
\end{figure}
\begin{equation}
a_1^{(2B)}=0.3145(1).
\end{equation}
It is a stable zero, so it should describe an observable stable stationary state
of a moduli field.
One possibility is to interpret
this as a strong coupling
at the lightest glueball mass scale. The lightest scalar glueball has spin $J^{PC}=0^{++}$
and is denoted by $gg^{0++}$ in the following. In our oscillator
interpretation we take
$B=gg^{0++}$,
$f_1=u$, $\bar{f_2}=\bar{u}$, thus
\begin{equation}
a_1^{(2B)}=\alpha_s (m_{gg^{0++}}+2m_u) \approx \alpha_s(m_{gg^{0++}}).
\label{glueball}
\end{equation}
The 2B dynamics then describes two bosons at the same time (two
gluons forming a glueball), similar to the 3B
dynamics, which described two fermions at the
same time (left and right handed). In both cases a spin 0 state is formed in total.
In lattice gauge
calculations including dynamical fermions the smallest scalar glueball
mass is estimated as $m_{gg^{0++}}=(1.74 \pm 0.07)$ GeV \cite{100} and
at this energy the running strong coupling constant is
experimentally measured to be $\alpha_s \approx 0.32$.
This clearly is consistent with the observed value
of $a_1^{(2B)}$.

\subsection{The 2A$\,^-$ and 2B$\,^-$ dynamics --- towards a Higgs mass prediction}
Two chaotic field theories are still remaining, namely those with
$N=2$ and antidiffusive coupling. The interaction energies $W(a)$
of the 2A$\,^-$ and 2B$\,^-$ chaotic fields are shown in Fig.~6
and 7. Let us now try to find a suitable physical interpretation
for the observed smallest stable zeros $a_1^{(2A^-)}=0.1758(1)$
and $a_1^{(2B^-)}=0.095370(1)$.
\begin{figure}
\epsfig{file=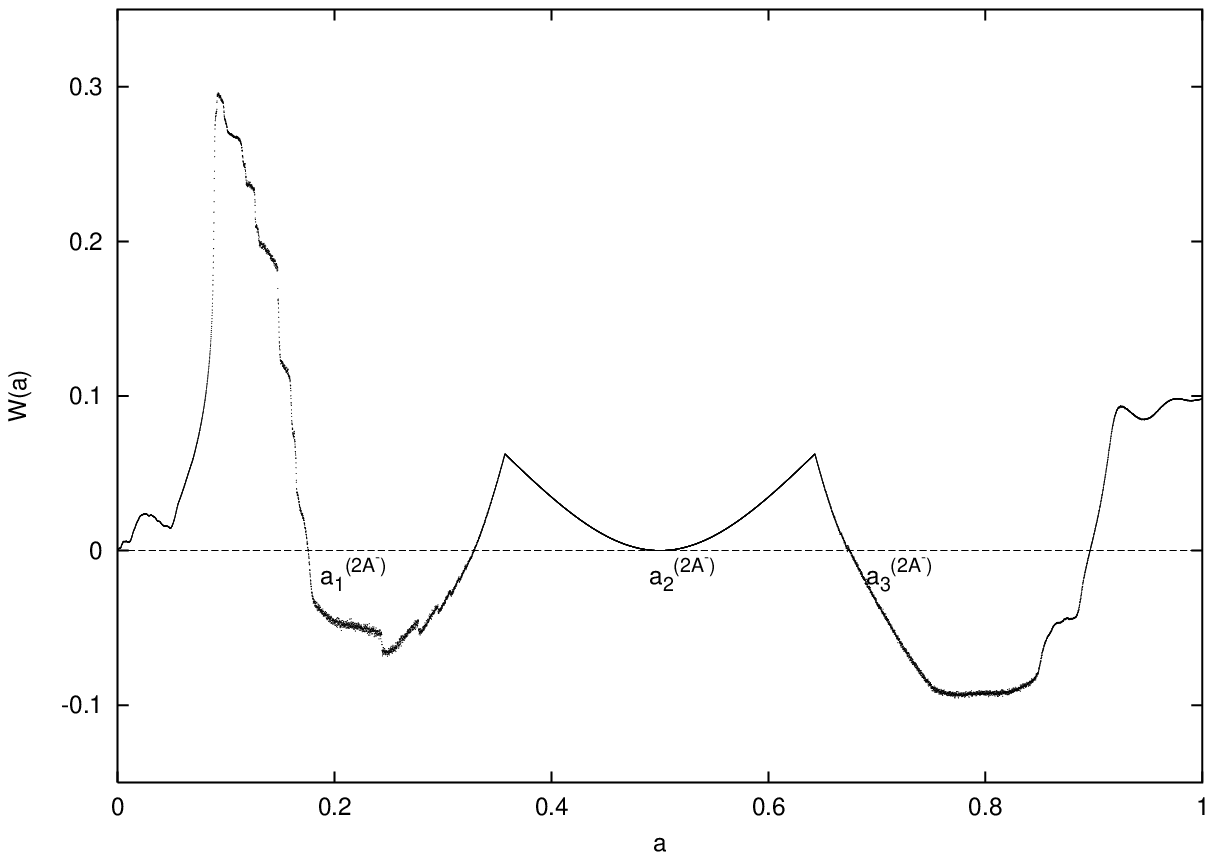, width=10cm, height=8cm}
\caption{Interaction energy of the 2$A^-$ dynamics.}
\end{figure}
\begin{figure}
\epsfig{file=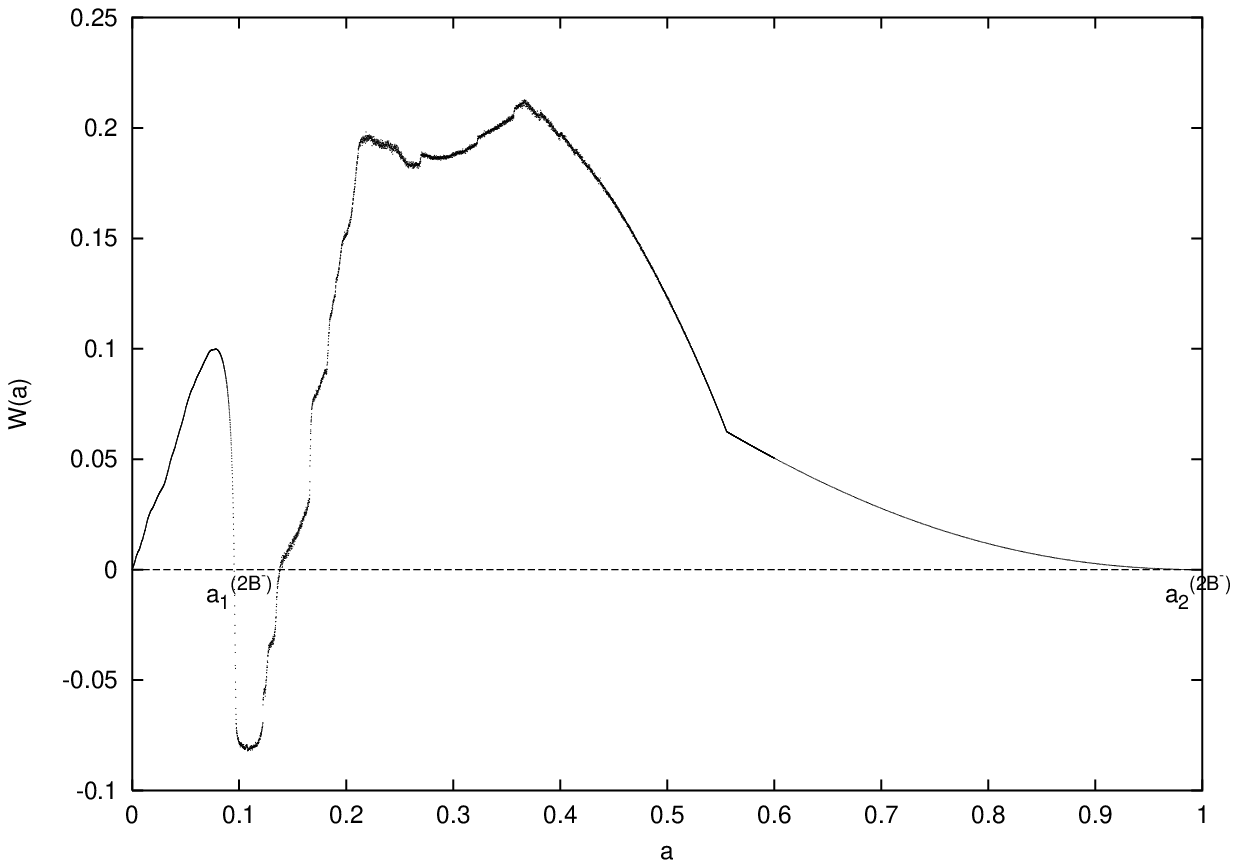, width=10cm, height=8cm}
\caption{Interaction energy of the 2$B^-$ dynamics.}
\end{figure}
Again let us be guided by discrete symmetry considerations when
attributing the various stationary moduli states to standard model
particles. We saw that the smallest stable zero of the 2A dynamics
describes a boson with non-zero spin (the lightest massive gauge
boson $W^\pm$) and the smallest stable zero of the 2B dynamics a
boson without spin (the lightest scalar glueball). Thus it seems
reasonable to assume that the smallest stable zero of the
2A$\,^-$ gives us information on yet another bosonic particle
with non-zero spin, possibly the lightest glueball with spin
$J^{PC}=2^{++}$, and the smallest zero of the 2B$\,^-$ dynamics
on yet another bosonic particle with spin 0, possibly the
lightest Higgs boson.

Let us start with $a_1^{(2A^-)}$. Our physical interpretation of
this stationary moduli field state in the sense of
eq.~(\ref{katharina}) is $B=gg^{2++}, f_1=q, \bar{f}_2=\bar{q}$,
where $q, \bar{q}$ are suitable quarks. From the strong coupling
interpretation
\begin{equation}
a_1^{(2A^-)}=\alpha_s (E_1)
\end{equation}
the energy $E_1=m_{gg^{2++}}+2m_q$ can again be determined from
the usual QCD formula, evolving our previously determined
$\alpha_s(m_W)$ to lower energies. One obtains $E_1\approx 10.45$
GeV. In lattice gauge calculations the mass of the lightest
$2^{++}$ glueball is estimated as being roughly $2$ GeV. We thus
get the correct order of magnitude of the $2^{++}$ glueball mass
if we assume that the quarks in eq.~(\ref{katharina}) are bottom
quarks. With this interpretation a glueball mass
$m_{gg2^{++}}=(10.45-2\cdot 4.23)$ GeV =2.0 GeV is predicted,
using $m_b=4.23$ GeV.

Next, let us consider $a_1^{(2B^-)}$. Having already obtained
information on the $W$ boson in section 6.3, it is likely
that there are also moduli fields that encode the Higgs boson.
Our physical interpretation is $B=H, f_1=q, \bar{f}_2=\bar{q}$,
where $H$ is the lightest Higgs boson and $q,\bar{q}$ are suitable
quarks. The strong coupling interpretation
\begin{equation}
a_1^{(2B^-)}=\alpha_s (E_2)
\end{equation}
yields $E_2=483.4(3)$ GeV =$m_H+2m_q$. However, experimental and
theoretical arguments imply that the Higgs mass should be in the
region 100...200 GeV. Hence we only obtain a consistent value for
the Higgs mass if we assume that the quarks involved are $t$
quarks. This is similar to the zero $a_1^{(2A^-)}$, where the
quarks involved were also heavy quarks. Generally, chaotic fields
with antidiffusive couplings seem to encode information on heavy
quarks rather than light ones.

From the self energy of the 2B dynamics, one can obtain quite a
precise prediction of the free top mass, namely $m_t=164.5(2)$
GeV, corresponding to a top pole mass of 174.4(3) GeV (see
\cite{physicad, book} for more details). With this value the zero
$a_1^{(2B^-)}$ yields a Higgs mass prediction of
\begin{equation}
m_H=E_2-2m_t=154.4(5) GeV.
\end{equation}
This is a very precise prediction, the statistical error is very
small. But of course the main source of uncertainty is a
theoretical uncertainty, namely whether our harmonic oscillator
interpretation (in the sense of eq.~(\ref{katharina})) of the zero
$a_1^{(2B^-)}$ in terms of a Higgs boson and two top quarks is correct.
For example, assuming that a supersymmetric extension of the
standard model is correct, then the zero could also describe other
particle states whose masses add up to 483.4 GeV.

\section{New physics?}

A few stable zeros of interaction energies remain that cannot be
interpreted in the standard model context. The $3A$ dynamics has
yet another stable zero $a_3^{(3A)}=0.07318$. Interpreted as a
running strong coupling, this would correspond to an energy $E=
\frac{3}{2} (m_B +m_{f_1}+m_{\bar{f_2}}) \approx 7.85$ TeV. One
could speculate that this might describe a (rather large)
supersymmetric particle scale. There are a few further stable
zeros in the large coupling region: $a=0.9141$ and $a=1$ ($3A$),
$a=0.3496$ ($3B$), $a=0.675$ ($2A^-$) and $a=1$ ($2B^-$).
Possibly these stationary moduli states could describe
gravitational couplings at the Planck scale.

So far we only talked about the interaction energy of chaotic
fields. But one can also look at the self energy as given in
eq.~(\ref{39}) or (\ref{40}). Again a large number of local
minima are observed that can be associated with known standard
model interaction strengths, the energy $E$ being again given by
eq.~(\ref{katharina}). We don't have the space to describe all
the details here, but refer to \cite{physicad, book}. From local
minima of the self energy one can get some rather precise
predictions of fermion masses, in particular for the heavy
fermions $t,b,c,\tau$. The `landscape' generated by the self
energy has also minima that can be identified with Yukawa
couplings and gravitational couplings.

The lighter fermions are much more difficult to deal with in this
context. Since for very small couplings the self energy exhibits
scaling behaviour with log-periodic oscillations \cite{groote},
one is only able to give predictions of light fermion masses
modulo 2, and there are also some other theoretical ambiguities on
how to associate the various minima with the light particles. For
an early attempt to predict neutrino masses, see \cite{physicad}.
Clearly more work is needed to eliminate the ambiguities for light
fermions.

Proceeding to higher energies (by investigating the chaotic
dynamics for coupling constants $a$ that coincide with running
standard model coupling constants couplings $\alpha_1,
\alpha_2,\alpha_3$ at large energies), and using again the self
energy as the relevant observable, one can also look for local
minima that potentially could describe supersymmetric particles
with masses in the TeV region. No such minima are found
\cite{book}. Generally the evidence for supersymmetric particles
in the TeV region from our chaotic fields is pretty meagre, to say
the least. One would expect to see lots of local minima in the
relevant coupling region if the conventional ideas on
supersymmetry breaking at the TeV scale are correct. The self
energies of the chaotic fields do not provide any evidence for
such particles: There are simply no minima in the relevant
coupling region, the only exception being perhaps the zero
$a_3^{(3A)}$ of the 3A interaction energy. However, minima {\em
are} found at the energy scale $10^{16}$ GeV and at the Planck
scale. See \cite{book} for more details. If future accelerator
experiments do not find any evidence for supersymmetric particles,
a theoretical reason could be that they do not fit into the
`landscape' generated by the chaotic fields.

Talking about new physics, new physics is of course represented
by the existence of the chaotic fields themselves.
For small couplings $a$, their equation of state is close to
$w=-1$ \cite{prd}, hence they can account for dark energy in the
universe. But for large couplings they can have an equation of
state close to $w \approx 0$
\cite{prd}. Thus, in principle at least, chaotic fields could
account for both, dark energy {\em and} dark matter in the
universe.

\section{Conclusion}

We have introduced chaotic scalar fields of as a model of vacuum
fluctuations in the dark energy sector. These chaotic fields were
used to generate potentials for moduli fields. We numerically
observe that minima of the potentials lead to realistic standard
model coupling strengths. The values of the fine structure
constant, of the weak mixing angle, and of the strong coupling at
the $W$ mass scale are obtained with high precision
and correspond to stable stationary values of moduli fields. Based
on additional discrete symmetry assumptions, a value of the Higgs
mass of 154 GeV is predicted from the chaotic field dynamics.
This prediction can be experimentally tested in the near future.

The theory described in this paper may be regarded as a `tip of an
iceberg'. It still needs to be embedded into a greater context. In
particular, its possible relation to supersymmetry breaking
mechanisms has to be clarified. However, one statement can be
made without any doubt: The nonlinear dynamics given by
eq.~(\ref{dyn}) appears to distinguish certain numerical values
of coupling constants $a$ that do coincide with known standard
model coupling constants with very high precision. A random
coincidence can really be excluded.
In this way the chaotic fields can help to select the `correct'
vacuum out of an enormous number of possibilities,
shaping the world around us in precisely
the way we know it.

\end{document}